# A New Underdetermined Framework for Sparse Estimation of Fault Location for Transmission Lines Using Limited Current Measurements

Guangxiao Zhang, *Member, IEEE*, Gaoxi Xiao, *Senior Member, IEEE*, Xinghua Liu, *Senior Member, IEEE*, Yan Xu, *Senior Member, IEEE*, and Peng Wang, *Fellow, IEEE*

*Abstract*— This letter proposes an alternative underdetermined framework for fault location that utilizes current measurements along with the branch-bus matrix, providing another option besides the traditional voltage-based methods. To enhance fault location accuracy in the presence of multiple outliers, the robust YALL1 algorithm is used to resist outlier interference and accurately recover the sparse vector, thereby pinpointing the fault precisely. The results on the IEEE 39-bus test system demonstrate the effectiveness and robustness of the proposed method.

## I. INTRODUCTION

Traditional fault location methods, such as single-ended, double-ended, or multi-ended approaches, are tailored for single transmission line and require significant equipment and infrastructure for deployment on each line across a large-scale power network. With the advent of advanced wide-area measurement and communication technologies, research interest has been sparked in wide-area fault location methods, including traveling wave-based and impedance-based approaches [1]. These methods utilize data from extensive geographic areas, collected by sparsely placed measurement devices, enabling precise fault location on transmission lines throughout power networks without requiring the installation of equipment and infrastructure on every line.

While traveling-wave-based methods offer higher precision, the rely on high sampling rates for data acquisition. The impedance-based wide-area fault location methods are broadly categorized into methods based on overdetermined systems and those based on underdetermined systems. Overdetermined systems-based methods are formulated using limited voltage/current measurements and typically employ the least squares algorithm for solving [2]. These methods require the creation of specific overdetermined equations for each transmission line and a comparison of results from all such equations to determine the faulted line and location. Such a process becomes particularly burdensome in transmission networks with many lines. Underdetermined system-based methods recast fault location in power networks as a sparse estimation problem, employing sparse recovery algorithms to identify variables, where non-zero elements indicate the faulted lines and are used to calculate the fault location [3].

However, to our knowledge, existing underdetermined systems-based wide-area fault location methods have been developed using voltage measurements, with no such methods based on current measurements. Considering that current signals also contain rich fault information, this letter proposes an alternative underdetermined framework for fault location that utilizes current measurements in conjunction with the branch-bus matrix, offering an additional option beyond the traditional voltage-based methods. Additionally, the robust YALL1 algorithm is employed to resist interference from multiple outliers in order to achieve precise fault location.

## II. PROPOSED METHOD

### A. Underdetermined System for Fault Location Using Current Measurements

For an $N$-bus transmission network, the voltage sag phasor at any bus for a fault at bus $k$ in the network can be expressed as:

$$\Delta V_p = Z_{pk}\Delta I_k \qquad (1)$$

where $\Delta V_p$ is voltage phasor difference between the positive-sequence pre- and post-fault voltage phasors at any bus $p$ ($p = 1, 2, …, N$). $\Delta I_k$ is the positive-sequence current injected at the faulted bus $k$ ($k = 1, 2, …, N$). $Z_{pk}$ is the positive-sequence $p$-$k$ element of the bus-impedance matrix.

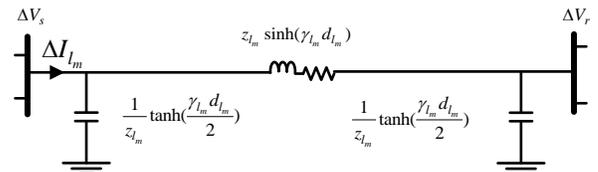

Fig.1. The distributed parameter model of a transmission line.

By applying Kirchhoff's voltage law (KVL) to the distributed parameter model of the transmission line shown in Fig. 1, the positive-sequence superimposed current of any line with terminals $s$ and $r$ can be obtained as:

$$\Delta I_{l_m} = \frac{\Delta V_s}{z_{l_m}}\tanh(\frac{\gamma_{l_m} d_{l_m}}{2}) + \frac{\Delta V_s - \Delta V_r}{z_{l_m}\sinh(\gamma_{l_m} d_{l_m})} \qquad (2)$$

where $\Delta I_{l_m}$ are the current differences calculated by the post- and pre-fault current phasors. The subscript $l_m$ ($m = 1, 2, …, M$) represents the serial number of the branches equipped with sensors. $z_{l_m}$, $\gamma_{l_m}$, and $d_{l_m}$ are the surge impedance, propagation constant, and length of line $l_m$, respectively. Referring to Eq. (1), we can rewrite $\Delta V_s$ and $\Delta V_r$ in Eq. (2) to obtain:

$$\Delta I_{l_m} = \beta_{l_m,k}\Delta I_k \qquad (3)$$

This work was supported by the Future Resilient Systems (FRS-II) Project at the Singapore-ETH Centre (SEC), which was funded by the National Research Foundation of Singapore (NRF) under its Campus for Research Excellence and Technological Enterprise (CREATE) program. (Corresponding author: *Gaoxi Xiao*.)

Guangxiao Zhang is with the institute of Catastrophe Risk Management, Nanyang Technological University, Singapore 639798, and with Future Resilient Systems, Singapore-ETH Centre, Singapore 138602. (e-mail: guangxiao.zhang@ntu.edu.sg).
Gaoxi Xiao, Yan Xu, and Peng Wang are with the School of Electrical and Electronic Engineering, Nanyang Technological University, Singapore 639798. (e-mail: egxxiao@ntu.edu.sg, xuyan@ntu.edu.sg, epwang@ntu.edu.sg).
Xinghua Liu is with the School of Electrical Engineering, Xi'an University of Technology, Xi'an 710048, China. (e-mail: liuxh@xaut.edu.cn).

with,

$$\beta_{l_m,k} = \frac{Z_{sk}}{z_{l_m}} \tanh(\frac{\gamma_{l_m} d_{l_m}}{2}) + \frac{Z_{sk} - Z_{rk}}{z_{l_m} \sinh(\gamma_{l_m} d_{l_m})}$$

Similarly, for a fault point $f$ along line $l_g$ with terminals $i$ and $j$ at distance $x$ from bus $i$, the Eq. (1) can be expressed as [2]:

$$\Delta V_p = Z_{ip}\Delta I_i + Z_{jp}\Delta I_j \qquad (4)$$

with,

$$\Delta I_i = \frac{\sinh(\gamma_{l_g} l_g (1-x))}{\sinh(\gamma_{l_g} l_g)} I_f, \ \Delta I_j = \frac{\sinh(\gamma_{l_g} l_g x)}{\sinh(\gamma_{l_g} l_g)} I_f$$

where $I_f$ is the positive-sequence fault current at fault point $f$.

Likewise, $\Delta I_{lm}$ in Eq. (2) can be rewritten in terms of $\Delta I_i$ and $\Delta I_j$ as follows.

$$\Delta I_{l_m} = \beta_{l_m,i}\Delta I_i + \beta_{l_m,j}\Delta I_j \qquad (5)$$

Accordingly, Eq. (5) can be extended into the following linear system of equations.

$$\begin{bmatrix} \Delta I_{l_1} \\ \Delta I_{l_2} \\ \vdots \\ \Delta I_{l_m} \\ \vdots \\ \Delta I_{l_M} \end{bmatrix}_{M \times 1} = \begin{bmatrix} \beta_{l_1,1} & \cdots & \beta_{l_1,i} & \beta_{l_1,j} & \cdots & \beta_{l_1,N} \\ \beta_{l_2,1} & \cdots & \beta_{l_2,i} & \beta_{l_2,j} & \cdots & \beta_{l_2,N} \\ \vdots & & \vdots & \vdots & & \vdots \\ \beta_{l_m,1} & \cdots & \beta_{l_m,i} & \beta_{l_m,j} & \cdots & \beta_{l_m,N} \\ \vdots & & \vdots & \vdots & & \vdots \\ \beta_{l_M,1} & \cdots & \beta_{l_M,i} & \beta_{l_M,j} & \cdots & \beta_{l_M,N} \end{bmatrix}_{M \times N} \begin{bmatrix} 0 \\ \vdots \\ \Delta I_i \\ \Delta I_j \\ \vdots \\ 0 \end{bmatrix}_{N \times 1} \qquad (6)$$

Let $\vec{y} = [\Delta I_{l1}, \Delta I_{l2}, \ldots, \Delta I_{lm}, \ldots, \Delta I_{lM}]^T$ denote the observable complex current measurements vector and $\vec{\theta} = [0, \ldots, \Delta I_i, \Delta I_j, \ldots, 0]^T$ the unknown complex fault injection vector, where the nonzero elements only exist in the terminal nodes of the faulted line. Equations (2) and (6) can therefore be rewritten as:

$$\vec{y} = \vec{\beta}\vec{\theta} \qquad (7)$$

where $\vec{\beta} \in \mathbb{C}^{M \times N}$ is the complex branch-bus matrix where each row corresponds to one of the $M$ observable branches and each column to one of the $N$ buses. Note that when only a limited number of branches are equipped with sensors, $M \ll N$, making Eq. (7) an underdetermined system of which the number of equations is fewer than the number of variables. The complex equation can be converted to real values for use with sparse recovery algorithms [4].

$$y = \beta\theta + e \qquad (8)$$

with,

$$y = \begin{bmatrix} \text{Re}(\vec{y}) \\ \text{Im}(\vec{y}) \end{bmatrix}, \ y = \begin{bmatrix} \text{Re}(\vec{\beta}) & -\text{Im}(\vec{\beta}) \\ \text{Im}(\vec{\beta}) & \text{Re}(\vec{\beta}) \end{bmatrix}, \ \theta = \begin{bmatrix} \text{Re}(\vec{\theta}) \\ \text{Im}(\vec{\theta}) \end{bmatrix}$$

where Re ($\cdot$) represents the real part of a complex vector, and Im ($\cdot$) denotes the imaginary part. $e \in \mathbb{R}^{2N \times 1}$ is additive measurement noise.

Once the fault injection vector $\theta$ is recovered, the faulted line $l_g$ can be indicated in the transmission networks. Thus, the exact fault location $x$ can be calculated as:

$$x = \ln[\frac{e^{\gamma_{l_g} d_{l_g}} + \Delta I_i / \Delta I_j}{e^{-\gamma_{l_g} d_{l_g}} + \Delta I_i / \Delta I_j}]/2\gamma_{l_g} d_{l_g} \qquad (9)$$

### B. Robust Sparse Recovery of Fault Location Against Outliers

Equation (8) represents a typical compressed sensing problem, where the unknown variable $\theta$ with inherent sparsity can be efficiently recovered from a limited number of measurements. Due to the potential outliers caused by factors like sensor malfunctions, erroneous measurements and cyber threats, the $L_2$-norm model for data fitting proves vulnerable to such bad data. Therefore, the $L_1$-norm model is employed to derive a more robust formulation as follows:

$$\underset{\theta}{\text{minimize}} \ \{\|\beta\theta - y\|_1 + \lambda\|\theta\|_1\} \qquad (10)$$

where $\lambda$ is the regularization parameter that controls the sparsity of the solution, and the term $\|\cdot\|_1$ denotes the $L_1$-norm.

The YALL1 algorithm, which utilizes the Alternating Direction Method of Multipliers (ADMM) framework, iteratively solves the $L_1$-norm minimization problem [5]. The solution involves updating three sets of variables: the auxiliary variable $v$, the sparse signal $\theta$, and the multiplier $w$. These updates proceed as follows:

1) Auxiliary Variable Update ($v^{k+1}$): The auxiliary variable aims to approximate the absolute discrepancies between the predicted and actual measurements. The update rule is:

$$v^{k+1} = \text{sign}(a) \cdot \max(|a| - \frac{\rho}{\lambda}, 0) \qquad (11)$$

with,

$$a = \beta\theta^k - y - \frac{w^k}{\rho}$$

where $v$ is the auxiliary variable for iterative updates. The index $k$ represents the iteration number, increasing from 0 until convergence. The parameter $\rho$ is the penalty term, and $w$ is the multiplier used in the updates.

2) Sparse Signal Update ($\theta^{k+1}$): The update of the sparse signal $\theta$ considers the regularization term $\lambda\|\theta\|_1$. The update rule is given by:

$$\theta^{k+1} = \text{sign}(b) \cdot \max(|b| - \frac{\tau}{\rho}, 0) \qquad (12)$$

with,

$$b = \theta^k - \tau[\beta^T(a - v^{k+1})]$$

where $\tau$ is a step size parameter in the update of $\theta$.

3) Multiplier Update ($w^{k+1}$): The multiplier reflects the dual ascent step in the ADMM framework, promoting constraint adherence. The update rule is:

$$w^{k+1} = w^k - \rho(\beta\theta^{k+1} - y - v^{k+1}) \qquad (13)$$

The algorithm is terminated if the change in $\theta^{k+1}$ and $v^{k+1}$ between consecutive iterations falls below a threshold.

### III. CASE STUDY

The performance of the proposed method is evaluated using PSCAD/EMTDC on the IEEE 39-bus test systems, as shown in Fig. 1. A limited number of PMUs were installed at branches 3-18, 5-6, 8-9, 11-6, 14-13, 16-21, 19-16, 23-22, 25-2, 27-26, 29-28, and 39-1 to obtain current measurements.

Table I presents the results of fault location accuracy from 1000 simulation runs for various types of faults on the transmission lines of the network. These fault types include three-phase-to-ground (TPG), double-line-to-ground (DLG), line-to-line (LL), and single-line-to-ground (SLG) faults, each occurring at different locations $x$ and with varying fault resistances $R_f$. During the simulations, the measurements were adjusted to include ±1% random errors. Table I shows that the average fault location estimation error consistently remains below 0.5% across all fault scenarios tested, demonstrating the accuracy of the proposed method in locating faults.

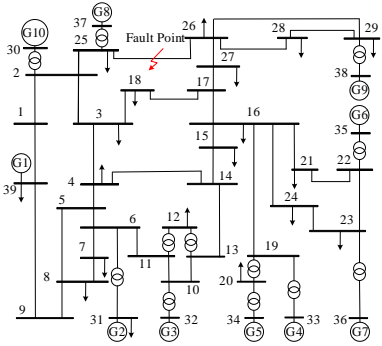

Fig. 2. IEEE 39-bus test systems.

TABLE I
ABSOLUTE PERCENTAGE ERROR IN FAULT LOCATION ACROSS DIFFERENT FAULT SCENARIOS

| Line | Type | $R_f$ (Ω) | Fault Distance $x$ (p.u.) | | | |
|---|---|---|---|---|---|---|
| | | | 0.1 | 0.3 | 0.5 | 0.9 |
| 4-5 | TPG | 0.1 | 0.1053 | 0.1173 | 0.1561 | 0.2050 |
| | | 100 | 0.1764 | 0.1715 | 0.1489 | 0.2226 |
| | DLG | 0.1 | 0.1035 | 0.1187 | 0.1497 | 0.1920 |
| | | 100 | 0.0620 | 0.1094 | 0.1395 | 0.1983 |
| | LL | 0.1 | 0.0831 | 0.1394 | 0.1444 | 0.2047 |
| | | 100 | 0.1207 | 0.0808 | 0.1487 | 0.3352 |
| | SLG | 0.1 | 0.0586 | 0.1477 | 0.1347 | 0.1965 |
| | | 100 | 0.0956 | 0.1500 | 0.0913 | 0.2382 |
| 21-22 | TPG | 0.1 | 0.0323 | 0.0469 | 0.0432 | 0.0355 |
| | | 100 | 0.1554 | 0.2162 | 0.0896 | 0.1060 |
| | DLG | 0.1 | 0.0129 | 0.0419 | 0.0347 | 0.0252 |
| | | 100 | 0.0260 | 0.0336 | 0.0338 | 0.0214 |
| | LL | 0.1 | 0.1395 | 0.0430 | 0.0506 | 0.0552 |
| | | 100 | 0.2635 | 0.1144 | 0.0671 | 0.0903 |
| | SLG | 0.1 | 0.0753 | 0.0046 | 0.0242 | 0.0875 |
| | | 100 | 0.2322 | 0.1319 | 0.0539 | 0.0511 |
| 25-26 | TPG | 0.1 | 0.0653 | 0.0855 | 0.0868 | 0.0921 |
| | | 100 | 0.0563 | 0.1357 | 0.0872 | 0.1084 |
| | DLG | 0.1 | 0.0740 | 0.0935 | 0.0849 | 0.0939 |
| | | 100 | 0.0485 | 0.0614 | 0.0983 | 0.1119 |
| | LL | 0.1 | 0.0734 | 0.1154 | 0.1138 | 0.0907 |
| | | 100 | 0.1819 | 0.1956 | 0.2034 | 0.1139 |
| | SLG | 0.1 | 0.0206 | 0.1019 | 0.0497 | 0.2350 |
| | | 100 | 0.1392 | 0.1930 | 0.1030 | 0.2617 |

To demonstrate the robustness of the proposed method, an TPG fault at line 21–22 with a fault resistance of 0.1 Ω located at 10% of the line length was simulated. The measurements were adjusted to include ±1% random errors, and 20% of the data points were affected by outliers, with both the real and imaginary parts of the clean and corrupted measurement data illustrated in Fig. 3. Figure 4 provides a comparative performance analysis, including the recovered vectors and the corresponding relative errors for the Lasso algorithm [3], Huber-FISTA algorithm [6], and the YALL1 algorithm employed in this study. From Figs. 4 (a) and (d), it is evident that the Lasso-based method fails to recover the target vector, with a normalized error of 0.562. The Huber-FISTA method achieves better recovery with a normalized error of 0.015, as shown in Figs. 4 (b) and (e). The YALL1 algorithm utilized in this paper closely approximates the actual result, with a normalized error of just 0.001, as depicted in Figs. 4 (c) and (f). Based on the vectors recovered by the Lasso algorithm, Huber-FISTA algorithm, and the YALL1 algorithm used in this study, the calculated fault locations are 0.3255, 0.0912, and 0.1004, respectively, with corresponding location errors of 22.5466%, 0.8845%, and 0.0443%. This demonstrates that the Lasso algorithm, lacking robustness, yields significantly biased estimations under outlier interference. Although Huber-FISTA has some tolerance for outliers, its performance in handling large or severe outliers is still inferior to YALL1, which more effectively reduces the impact of outliers by optimizing the $L_1$-norm.

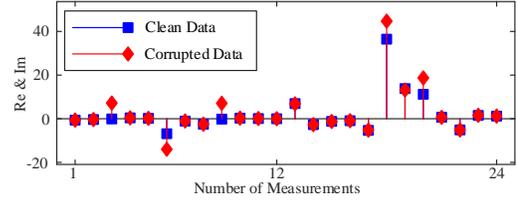

Fig. 3. Real and imaginary parts of measurement data with clean data and corrupted data.

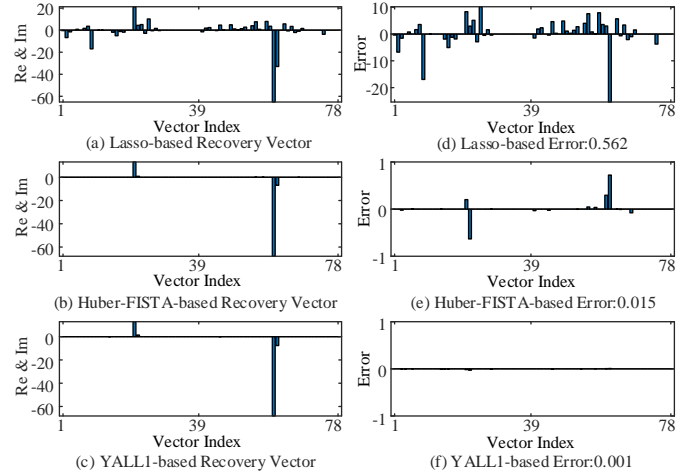

Fig. 4. Comparison of recovery vectors and errors for Lasso-based, Huber-FISTA-based, and YALL1-based algorithms under outlier interference.

IV. CONCLUSION

This letter proposed a novel underdetermined framework for fault location that utilizes sparse current measurements instead of traditional voltage measurements, in conjunction with the positive-sequence branch-bus matrix. To address the impact of multiple outliers on the accuracy of fault location, a robust YALL1 algorithm was employed to mitigate interference from outliers and achieve precise recovery of the vector, thereby pinpointing the fault accurately. The effectiveness of the proposed method was validated under various fault scenarios and in the presence of outlier interference.